\documentstyle[12pt]{article}
\begin{document}
\renewcommand{\baselinestretch}{1.37}
\small\normalsize
\renewcommand{\theequation}{\arabic{section}.\arabic{equation}}
\renewcommand{\thesection}{\arabic{section}.}
\language0

\begin{flushright}
{\sc BUHEP}-96-1\\January 1996
\end{flushright}

\vspace*{1.5cm}

\begin{center}

{\Large \bf Order parameters and boundary effects\\in U(1) lattice 
gauge theory}

\vspace*{0.8cm}

{\bf Werner Kerler$^a$, Claudio Rebbi$^b$, Andreas Weber$^a$}

\vspace*{0.3cm}

{\sl $^a$ Fachbereich Physik, Universit\"at Marburg, D-35032 Marburg, 
Germany\\ 
$^b$ Department of Physics, Boston University, Boston, MA 02215, USA}
\hspace*{3.6mm}
       
\end{center}

\vspace*{1.5cm}

\begin{abstract}
We show that, independently of the boundary conditions, the two phases
of the 4-dimensional compact U(1) lattice gauge theory can be
characterized by the presence or absence of an ``infinite'' current
network, with an appropriate definition of ``infinite'' for the
various types of boundary conditions imposed on the finite lattice.
The probability for the occurrence of an ``infinite'' network takes
values 0 or 1 in the cold and hot phase, respectively.  It thus
constitutes a very efficient order parameter, which allows one to
determine the transition region at low computational cost. In
addition, for open and fixed boundary conditions we address the
question of the impact of inhomogeneities and give examples of the
reappearance of an energy gap already at moderate lattice sizes.
\end{abstract}

\newpage

\section{Introduction}

\hspace{3mm}
Recently, using periodic boundary conditions, we have shown \cite{krw95}
that the phases of the 4-dimensional compact U(1) lattice gauge theory
are unambiguously characterized by the topological properties of
monopole current lines as well as by those of minimal Dirac sheets.  Our
topological analysis, based on mappings preserving homotopy, relied
heavily on the chosen type of boundary conditions. Nevertheless we
conjectured that the phases are more generally characterized by the
presence or absence of an infinite structure, where a suitable definition, 
dependent on the boundary conditions, must be given for the meaning 
of ``infinite'' in presence of a finite lattice.

A major motivation for this work has been the verification of this
conjecture.  For this purpose we have studied systems with periodic, open
and fixed boundary conditions. For periodic boundary conditions our
results establish that the topological criterion is a much better
estimator of the phases than the extension of the largest monopole
network, which has also been proposed as an order parameter.  For open
and fixed boundary conditions we give an appropriate definition of
``infinite'' network and demonstrate that the phases are again determined
by the presence or absence of an infinite network.

A second motivation for the present work is related to the fact that the
widely accepted first order nature of the phase transition has recently
been put into question.  Indeed the authors of Ref.~\cite{ln94}, modeling
the U(1) system on the surface of a 5-dimensional cube rather than on the
usual 4-dimensional torus, found no evidence of an energy gap and
suggested that on a manifold with trivial homotopy group the energy gap
disappears. This has initiated other investigations where fixed boundary
conditions \cite{bf94}, suppression of monopoles at the boundaries
\cite{lbbs94} and open boundary conditions \cite{krw95} have been used to
study the effects of the boundary conditions.

In order to draw conclusions on such issues it is crucial to gain a
proper understanding of the effects of the inhomogeneity in the Euclidean
space distribution determined by the choice of non-periodic boundary
conditions.  Therefore, for the cases of open and fixed boundary
conditions we address this question as well. By suitable examples we show
that the gap reappears already on lattices of moderate sizes so that its
disappearance for certain types of boundary conditions is not absolute.

It has been known for some time \cite{ms82,h82} that for fixed and open
boundary conditions the Wilson loop averages provide upper and lower
bounds, respectively, on those obtained with periodic boundary
conditions. Here we consider the effects of the inhomogeneities on the
energy distribution: we will show that if one measures the energy
distribution within shells at fixed distance from the boundary, large
shifts of the peaks occur when one moves from shell to shell.

\section{Order parameters}

\hspace{3mm}
In this section we show that for periodic, open and fixed boundary
conditions the presence or absence of an infinite network of monopole
currents can indeed be taken as a criterion to distinguish between the
phases, validating our generalized conjecture.  We will give, of course,
the appropriate definitions of ``infinite'' for the three types of
boundary conditions.  It will be seen that this criterion provides an
order parameter which has the virtue of taking the values 1 and 0 in the
hot and cold phases, respectively, and is thus superior to the one
\cite{hw89} based on the relative size of the largest network
$n_{\mbox{\scriptsize max}}/ n_{\mbox{\scriptsize tot}}$.  Our order
parameter allows one to determine the transition region at very low
computational cost because the analysis of a single equilibrated
configuration is already sufficient. One should keep in mind that
the characterization of phases discussed here and the question of the
order of the transition are distinct issues.

For periodic boundary conditions, which give the lattice the connectivity
of a torus $\mbox{\bf T}^4$, we define ``infinite'' by ``topologically
nontrivial in all directions''. While for individual loops the
topological characterization is straightforward, for the dense networks
of monopole currents which occur in reality it requires the more
sophisticated analysis which we developed and described in
Refs.~\cite{krw94,krw95}.  Figure~1 gives our results for
$P_{\mbox{\scriptsize net}}$, i.e.~the probability for the occurrence of
a network which is nontrivial in all directions, as a function of
$\beta$.  It is obvious that this is an order parameter which takes
values exactly equal to 0 and 1 for the cold and hot phase, respectively,
and provides a better distinction between the phases than the ratio of
the size of the largest monopole network to the total number of monopoles,
$n_{\mbox{\scriptsize max}}/ n_{\mbox{\scriptsize tot}}$.

In view of various suggestions which have been formulated in the past, we
observe that a criterion based simply on the extension of the largest
network, rather than on the topological characterization we mentioned
above, is not adequate for periodic boundary conditions.  The meaning of
``infinite'' in this case would be that for each direction the
one-dimensional projection of the network covers the full extension of
the lattice. In a test performed at the phase transition point on a $16^4$
lattice, where the peaks of the energy distribution are well separated,
in the cold phase we have obtained the value 0.069(16) for this order
parameter (i.e.~the probability that the projection of the current
network over all four axes spans the whole lattice extent), as opposed to
0.000 for $P_{\mbox{\scriptsize net}}$ considered above. Thus it is
indeed neither extension nor size but rather the topology inherent in our
criterion which is relevant in the case of periodic boundary conditions.

For open boundary conditions the lattice is no longer self-dual. On the
dual lattice the current lines, i.e.~the lines formed by the links
carrying nonzero current, may end at the boundaries. In this case we
define ``infinite'' as ``touching the boundaries in all
directions''. Figure~1 shows that the probability $P_{\mbox{\scriptsize
net}}$ of getting the ``infinite'' network thus defined provides again
an order parameter which takes values 0 and 1 in the cold and hot phase,
respectively, and compares favorably with $n_{\mbox{\scriptsize max}}/
n_{\mbox{\scriptsize tot}}$.

Fixed boundary conditions are obtained by setting all gauge variables $U$
to $1$ at the boundary. The surface of the lattice is then made of
3-dimensional cubes with all their link variables equal to $1$.  On the
dual lattice these cubes correspond to links with vanishing monopole
current. Therefore, the subset on the dual lattice accessible to current
lines is a lattice with open boundary conditions (while the original
lattice with fixed boundary conditions is homeomorphic to the sphere
$\mbox{\bf S}^4$). Now we define ``infinite'' as ``reaching the
boundaries in all directions''. As can be seen from Figure~1 the
corresponding probability $P_{\mbox{\scriptsize net}}$ takes again values
0 and 1 in the two phases, and provides a much sharper distinction
between them than $n_{\mbox{\scriptsize max}}/ n_{\mbox{\scriptsize
tot}}$.

We have checked for all the boundary conditions considered that our
characterization of the phases by current networks not only works with
Wilson's action but also with the extension obtained by adding a monopole
term with positive or negative coupling, i.e.~with the action
\cite{bss85},
\begin{equation}
S=\beta \sum_{\mu>\nu,x} (1-\cos \Theta_{\mu\nu,x})+
\lambda \sum_{\rho,x} |M_{\rho,x}| \; , 
\end{equation}
where $M_{\rho,x} = \epsilon_{\rho\sigma\mu\nu}
(\bar{\Theta}_{\mu\nu,x+\sigma}-\bar{\Theta}_{\mu\nu,x})/4\pi$ and the
physical flux $\bar{\Theta}_{\mu\nu,x}\in [-\pi,\pi)$ is given by
$\Theta_{\mu\nu,x}=\bar{\Theta}_{\mu\nu,x}+2\pi n_{\mu\nu,x}$
\cite{dt80}. We have verified that our criterion gives good results
irrespective of lattice size (within the range of sizes available to us):
in particular, it works with a $16^4$ lattice as well as with an $8^4$
lattice.

Comparing the results for different boundary conditions presented in
Figure~1 it is apparent that for open and fixed boundary conditions the
transition region is much larger than for periodic boundary conditions.
The observed enlargement of the transition region reflects the
considerably larger finite-size effects present for open and fixed
boundary conditions.

In Ref.~\cite{krw95} we have shown that for periodic boundary conditions,
instead of considering current networks, the phases can be characterized
equally well by the topological properties of minimal Dirac sheets
(obtained by minimizing the the number of Dirac plaquettes by a
simulated-annealing gauge-fixing procedure). We found there that this
holds true not only for the Wilson action but also if we add to it a
monopole term with positive or negative coupling.

In this study we have performed a Dirac-sheet analysis for fixed boundary
conditions, too. For the Wilson action it turned out that the
characterization by minimal Dirac sheets works again, though with a
transition region somewhat more extended than the one obtained with the
network criterion.  With a monopole term, however, and for very large
negative couplings (e.g.~$\lambda=-2.0$), corresponding to large monopole
density, infinite Dirac sheets no longer form.  Thus the criterion
based current networks remains the only available criterion for the
characterization of the phases.

\section{Energy distributions}

\hspace{3mm}
In our simulations we have determined the plaquette energy distributions
$P(E)$ (where $E=(1/6L^4)\sum_{\mu>\nu,x} (1-\cos \Theta_{\mu\nu,x})$) as
well as the monopole density distributions. In the following we
illustrate only our results for $P(E)$ because the corresponding results
for the monopole density are very similar. We present data taken at the
phase transition, i.e.~at the value $\beta=\beta_c$ where the specific
heat is at a maximum, or very close to it.  We reproduce in Table 1 the
values which we have obtained for $\beta_c$ in correspondence to the
various lattice sizes and boundary conditions which we have used.

Figure~2 shows $P(E)$ obtained on a $16^4$ lattice. For periodic boundary
conditions the distribution exhibits a gap with well separated peaks,
while for open and fixed boundary conditions there is no gap. The shifts
of the averages are in agreement with the bounds given in
Ref.~\cite{ms82} (in the comparison of the two sets of results one should
remember that plaquette energy and Wilson loop factors are defined with
opposite signs).

It should be mentioned here that although on a $16^4$ lattice with
periodic boundary conditions the two peaks of the energy distribution are
separated by a substantial gap, the algorithm we introduced in
Ref.~\cite{krw95a} makes it still possible to perform simulations that
span both peaks.

With periodic boundary conditions no energy gap, indicative of a first
order transition, is seen on a $4^4$ lattice, which is clearly a
consequence of finite-size effects. One begins to observe a gap on an
$8^4$ lattice, with some overlap between the peaks, while the gap becomes
quite pronounced on a $16^4$ lattice, with large separation between the
peaks.  With the generalized action (2.1), for negative $\lambda$, where
the phase transition gets stronger, a gap occurs also on the $4^4$
lattice (it is visible for $\lambda=-1.0$ and the peaks are well
separated for $\lambda=-2.0$).

With open and fixed boundary conditions, there is no sign of a gap even
with a $16^4$ lattice, as is apparent from Figure~2.  The widths of the
peaks nevertheless remain relatively small which suggests that a more
complicated mechanism than simple smearing is at work.  The numerical
investigation of larger lattices is extremely demanding from the
computational point of view.  Thus the study of lattices large enough
that the effects of the finite size and of the inhomogeneity are no
longer important appears hardly feasible.

We have nevertheless been able to derive some information on whether the
disappearance of the gap is absolute or depends on the size and/or other
parameters of the system. In the following we demonstrate in two separate
cases that even with a system made inhomogeneous by the choice of
boundary conditions the gap can, in fact, be made to reappear.

In the first example we use fixed boundary conditions and the action
(2.1). From Figure~3 we see that with an $8^4$ lattice for sufficiently
large monopole density, i.e.~negative $\lambda$, a gap emerges again.
With open boundary conditions one finds very similar results.  This shows
that the effects of the inhomogeneity can be overcome by making the
transition stronger (which is quite analogous to the way in which
finite-size effects are overcome on the $4^4$ lattice with periodic
boundary conditions in the example mentioned above).

In the second example we use mixed boundary conditions, i.e.~fixed
boundary conditions only in direction 0 and periodic boundary conditions
in the other three directions. With these boundary conditions we observed
that there is no gap on a $8^4$ lattice. In Figure~4 we show what happens
if one considers a lattice of size $L_0\times 8^3$.  While for $L_0=8$,
as mentioned above, there is no gap, for $L_0=16$ a gap is seen to occur
again. Thus this simplified system demonstrates that the gap in the
energy distribution reappears even with non-periodic boundary conditions
when the system's size becomes sufficiently large.

In order to study the effects of inhomogeneities in detail we considered
shells (3-dimensional subsets) within the lattice and measured the
plaquette energy distribution on each of the shells separately.  We
number the shells by $s=1,\ldots,L/2$, where $L$ is the lattice
extension and $s=1$ corresponds to the outmost shell. A shell consists of
all lattice points where one of the coordinate is equal to $s-1$ or to
$L-s$.

Figure~5 shows the results obtained for each of the shells, separately,
in the same simulations as illustrated in Figure~2. It is manifest that
for open and fixed boundary conditions the distributions change
dramatically from shell to shell. We should note here that the $P(E)$ in
Figure~5 are normalized to the same constant for each individual shell.
Because of their larger weights, the outer shells are dominant in the sum
over the whole lattice.

In view of the large shifts apparent in Figure~5 one can hardly consider
results about the presence or absence of a gap to be reliable with these
types of boundary conditions unless one goes to much larger lattices
where the impact of boundaries decreases.

The directions of the shifts for open and fixed boundary conditions in
Figure~5 are in accordance with the lattice getting hotter and colder,
respectively, for outer shells. The case of fixed boundary conditions may
be described in terms of a picture \cite{h82} whereby, outside of the
lattice volume, all the dynamical degrees of freedom are
frozen. Similarly for open boundary conditions the finite lattice can be
considered as embedded into a ``hot environment''. It is conceivable that
this strong ``outside coupling'' is what prevents the occurrence of a gap
without affecting the widths of the peaks.

\section*{Acknowledgments}

\hspace{3mm}
This research was supported in part under DFG grants Ke 250/7-2 and 250/12-1 
and under DOE grant DE-FG02-91ER40676.
The computations were done on the CM5 at the Center for Computational
Science of Boston University and on the CM5 of the GMD at St.~Augustin.

\newpage
\renewcommand{\baselinestretch}{1.5}
\small\normalsize

\begin{center}
{\bf Table 1}\\
\vspace{5mm}

Phase transition points $\beta_c$\\
\vspace{5mm}
\begin{tabular}{|c|c|c|c|c|}
\hline
   boundaries &  lattice    & $\beta_c$  \\
\hline
   open     &    $8^4$      &   1.04(2)     \\
            &    $16^4$     &   1.023(3)    \\
   fixed    &    $8^4$      &   0.91(3)     \\
            &    $16^4$     &   0.981(3)    \\
  periodic  &    $8^4$      &   1.0075(1)   \\
            &    $16^4$     &   1.01084(5)  \\
  mixed     &    $8\times 8^3$    &   0.9901(3)   \\
            &    $16\times 8^3$   &   1.0051(3)   \\
\hline
\end{tabular}
\end{center}

\newpage
\renewcommand{\baselinestretch}{1.5}
\small\normalsize

\section*{Figure captions}

\begin{tabular}{rl} 
Fig.~1. & Order parameters $P_{\mbox{\scriptsize net}}$ and 
$n_{\mbox{\scriptsize max}}/ n_{\mbox{\scriptsize tot}}$
as functions of $\beta$ on $8^4$ lattices \\&
with periodic, open and fixed boundary conditions.\\

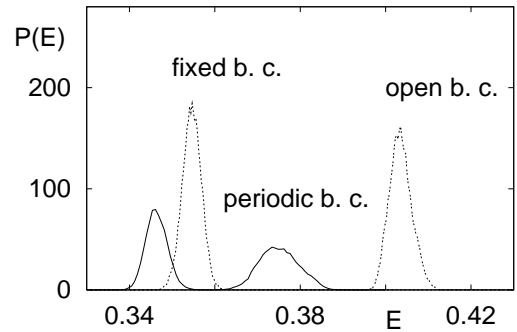
Fig.~2. & Probability distributions $P(E)$
on $16^4$ lattices with periodic, open and \\& 
fixed boundary conditions at $\beta=\beta_c$.\\

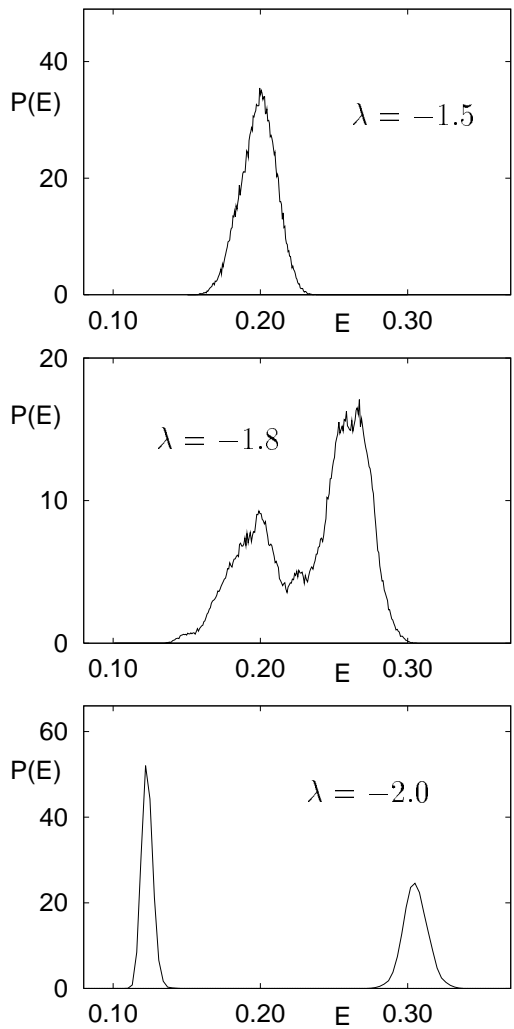
Fig.~3. & Probability distributions $P(E)$ on the $8^4$ lattice with fixed 
boundary \\& conditions and $\lambda=-1.5$, -1.8, -2.0 at $\beta=1.84$, 2.046, 
2.22, respectively.\\

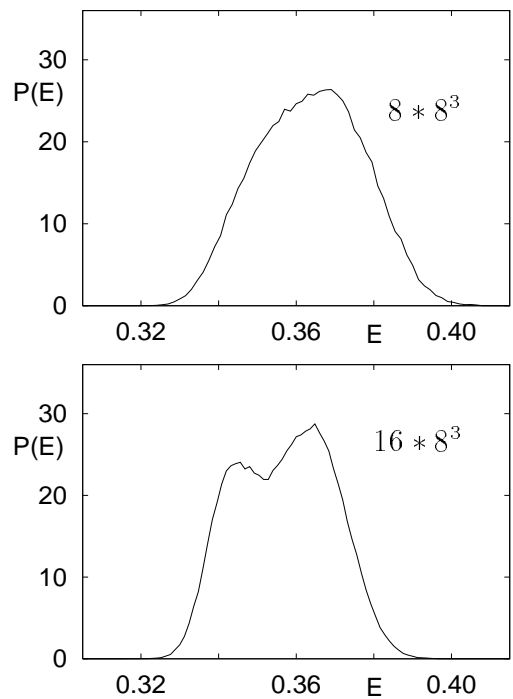
Fig.~4. & Probability distributions $P(E)$ on $8\times 8^3$ and 
$16\times 8^3$ lattices \\&
with mixed boundary conditions at $\beta=\beta_c$.\\

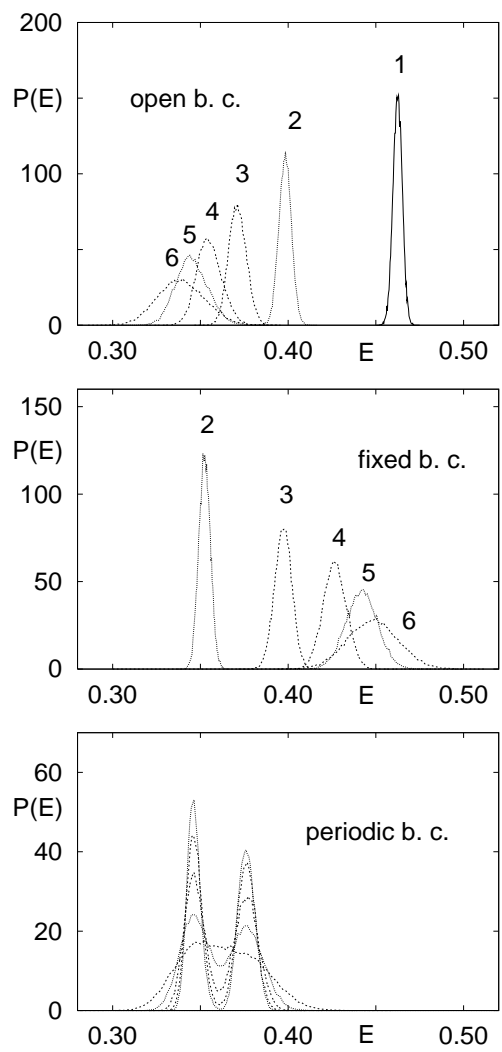
Fig.~5. & Probability distributions $P(E)$ for shells on $16^4$ lattices\\&
with periodic, fixed and open boundary conditions at $\beta=\beta_c$.\\

\end{tabular}

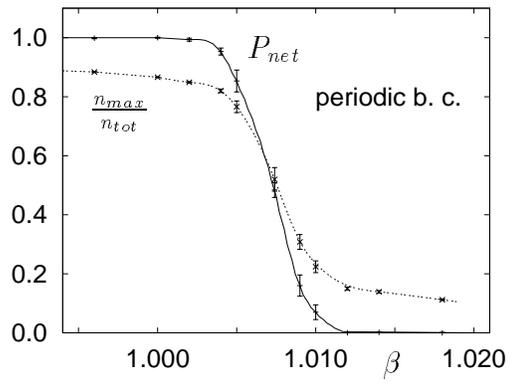
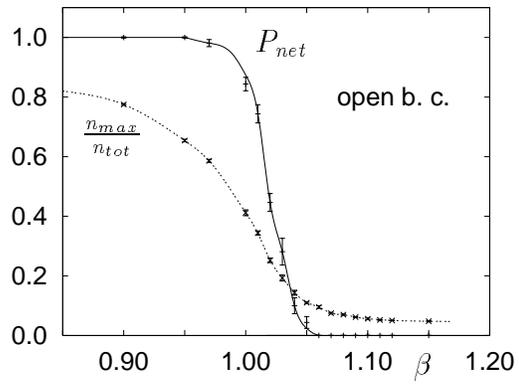
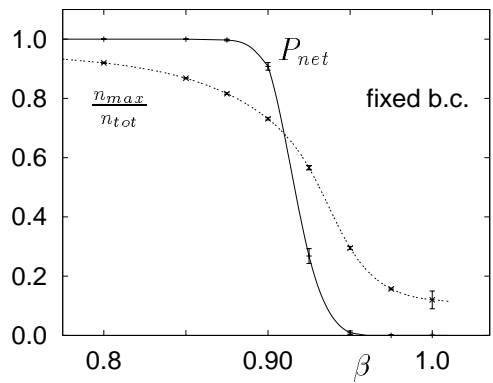
\end{document}